\documentclass[twocolumn]{aastex62}
\usepackage{amsmath}

\graphicspath{{./}{figures/}}

\usepackage{natbib}

\usepackage{amsmath}
\usepackage{amssymb}
\usepackage{xspace}
\usepackage{graphicx}

\newcommand{\ocy}{\omega_{\mathrm{cyc}}}

\newcommand{\lrphk}{\log_{10} R^{'}_\mathrm{HK}}

\def\s0#1#2{\mbox{\small{$ \frac{#1}{#2} $}}}
\def\0#1#2{\frac{#1}{#2}}

\newcommand{\pcyc}{P_{\mathrm{cyc}}}
\newcommand{\prot}{P_{\mathrm{rot}}}

\newcommand{\mh}{\left[\mbox{M/H}\right]}
\shorttitle{On the origin of the dichotomy in stellar activity cycles}
\shortauthors{Bonanno \& Corsaro}
\begin{document}
\title{On the origin of the dichotomy of stellar activity cycles}
\author{Alfio Bonanno}
\affiliation{INAF - Osservatorio Astrofisico di Catania, via S. Sofia 78, 95123-I, Italy}

\author{Enrico Corsaro}
\affiliation{INAF - Osservatorio Astrofisico di Catania, via S. Sofia 78, 95123-I, Italy}

\begin{abstract}
The presence of possible correlations between stellar rotation rate $\Omega$ and the frequency of
the activity cycle $\ocy$ is still much debated.  We implement a new Bayesian classification algorithm based on a simultaneous regression analysis of multiple scaling laws and we demonstrate the existence of two different scalings in the $\log_{10} \ocy$ -- $\log_{10} \Omega$ plane for an extended Mt.~Wilson sample of 67 stars. 
Thanks to metallicity measurements obtained from both ESA Gaia and high-resolution spectroscopy, we argue that the origin of this dichotomy is likely related to the chemical composition: stars whose magnetic cycle frequency increases with rotation rate are less metallic than stars whose magnetic cycle frequency decreases with stellar
rotation rates. On the contrary, no clear difference in chromospheric magnetic activity indicators characterizes the two branches. 
\end{abstract}
\keywords{stars: activity --- methods: statistical --- miscellaneous --- catalogs --- surveys}

\section{Introduction}
Despite the enormous progress made in recent years in the observations of stellar magnetism, its origin and  generation are still not completely understood.  It is believed that  the large-scale magnetic field observed in cool dwarfs  can be explained by the idea of ``cyclonic" turbulence \citep{1955ApJ...122..293P}: in a nutshell, 
a toroidal  field produced from a poloidal one by the shear ($\Omega$-effect) present in a stellar convection zone further  decays in helical turbulence  which ($\alpha$-effect) 
can then reinforce the dipole field \citep{1966ZNatA..21..369S}.
The magnetic field produced by the $\alpha\Omega$ dynamo mechanism is in general non-stationary 
and its late-time behavior is characterized by a power-law of the type
\begin{equation}
\ocy\propto \Omega^\nu
\label{eq:cyc_rot}
\end{equation}
where $\ocy \equiv 2 \pi / P_\mathrm{cyc}$ is the rate of the activity cycle, $\Omega \equiv 2 \pi / P_\mathrm{rot}$ is the rotation rate, and $\nu$ a scaling exponent. 
During the unstable phase, assuming $\Omega \propto \Omega' \ell $, $\Omega'$ being the shear and $\ell$ the characteristic scale of the turbulence, it turns out that
$\nu=1/2$ \citep{1976A&A....47..243S}, while in the opposite limit, in the saturated  phase, the cycle period is fixed by the characteristic turbulent diffusive time scale and $\nu=0$. 
For this reason it is reasonable to expect that in general $\nu$ will lay between these two limits, at least according to kinematic dynamo theory. At the non-linear level this picture can drastically change and it is not  clear if a simple scaling law like Eq.~(\ref{eq:cyc_rot}) holds at all
\citep{2018A&A...616A..72W,2019ApJ...880....6G,2021MNRAS.502.2565P}.

Pioneering  observational studies devoted to find possible correlation between $\ocy$ and $\Omega$ mostly agreed on $\nu\approx 1$ 
\citep{1978ApJ...226..379W,1984ApJ...287..769N,1996ApJ...460..848B} albeit the presence of multiple scaling laws (branches) has further complicated the discussion (\citealt{Saar99,2002ASPC..277..311S,2007vitense,2009AJ....138..312H,2016ApJ...826L...2M} but see also \citealt{2017LRSP...14....4B} for an extended review).

Stars can be classified in active or non-active according to the offset of two nearly identical scaling laws in the $\log_{10} (\ocy/\Omega)$ -- $\lrphk$ plane, the so-called RCRA (Ratio of Cycle over Rotation versus Activity) diagram. This classification plays a fundamental role in the analysis of stellar activity cycles because it determines the existence of possible correlations in the  $\pcyc$ -- $\prot$ plane  \citep{boro18}. 

Note that this has nothing to do with the existence of the Vaughan-Preston gap \citep{vp80}, a mild depletion in the number of stars around $\lrphk= -4.75$ caused by the weak bimodality of the distribution of $\lrphk$ in main-sequence stars \citep{2021A&A...646A..77G}.

The question has been further addressed in \cite{Olspert18} where a Gaussian classification algorithm has been used to determine the presence of multiple populations in the RCRA diagram. 
The problem with a direct clustering analysis in the RCRA diagram is the existence of possible correlations between $\ocy$ and $\Omega$ that would render the interpretation of the results ambiguous.

In this work we follow a different route. In Sect.~\ref{sec:bayes} we search for scaling laws of the type given by Eq.~(\ref{eq:cyc_rot}) in the $\log_{10}\ocy$ -- $\log_{10}\Omega$ plane
by considering randomly distributed couples of lines in this plane.
We thus employ a Bayesian classification algorithm based on a generalized likelihood describing a Gaussian mixture of two populations, so that 
the existence of branches is determined by the distance from the lines in this plane and not by the cluster center as in the approach followed by 
\cite{Olspert18}. We think that in the search for possible correlations between $P_\mathrm{cyc}$ and $P_\mathrm{rot}$ the $\log_{10}\ocy$ -- $\log_{10}\Omega$ plane has some advantages compared with the $P_\mathrm{cyc}$ -- $P_\mathrm{rot}$ plane used by some authors \citep[e.g.][]{2007vitense,2016ApJ...826L...2M,boro18} because it allows for a direct probe of a large class of power law dependencies as the one of  Eq.~(\ref{eq:cyc_rot}). Nonetheless, the data could be better reproduced by more complicated non-linear laws: one has to keep in mind that Eq.~(\ref{eq:cyc_rot}) is only a suggestion inspired by kinematic dynamo theory.

We present the data sample in Sect.~\ref{sec:data} and the data analysis in Sect.~\ref{sec:bayes}. We show our results in Sect.~\ref{sec:results} and discuss them in the light of available stellar properties in Sect.~\ref{sec:discussion}. Finally, we draw our conclusions on the origin of the two populations in Sect.~\ref{sec:conclusions}.

\section{Observations \& data}
\label{sec:data}
The analysis proposed in this work is based on measurements of rotation period, $P_\mathrm{rot}$ and magnetic activity cycle period, $P_\mathrm{cyc}$ that were published by \cite{boro18} (here after B18) and \cite{Olspert18} (hereafter O18). The adopted cycle periods are obtained from choromoshperic Ca H \& K measurements and are therefore excluding fast rotators, which are typically very young objects \citep[e.g.][]{Olah16,Lehtinen16,Distefano17}. The catalogs by B18 and O18 contain a substantial intersection of stars, for which multiple measurements are thus available. We therefore compiled the largest catalog of activity cycle measurements available to date 67 stars, including the Sun), by taking into account all stars in common between B18 and O18 (31 in total, including the Sun), as well as stars that belong to B18 only (14 stars) and to O18 only (22 stars). While for stars that are not common between the two catalogs the merging procedure is straightforward because only one set of measurements is available, for the stars falling in the intersection we applied the procedure described in the following.
O18 provides three estimates of $P_\mathrm{cyc}$, one based on a periodic model ($P_\mathrm{cyc,P}$), one on a harmonic model ($P_\mathrm{cyc,H}$), and another one on a quasi-periodic model ($P_\mathrm{cyc,QP}$). When possible we privilege the estimates obtained by means of a quasi-periodic model because i) it is the most commonly available and ii) it provides a more accurate fit to the time-series thanks to its higher complexity. For obtaining the final estimates of $P_\mathrm{cyc}$ we computed a weighted mean between the estimates from O18 and those from B18, i.e. taking into account the uncertainty in $P_\mathrm{cyc}$, except for the stars that we discuss below. HD~100180 has no $P_\mathrm{cyc}$ measurements from O18 so we considered the one from B18. HD~101501 was removed because the $P_\mathrm{cyc}$ measurements between the two catalogs deviate by more than 30\,\%. HD~149661 has only $P_\mathrm{cyc,P}$ from O18, while B18 provides two different estimates. Here we considered as a final $P_\mathrm{cyc}$ the weighted mean between O18 and the estimate from B18 that is closer to that of O18. HD~156026 has $P_\mathrm{cyc,H}$ and $P_\mathrm{cyc,P}$ from O18 that are almost identical, and an estimate from B18 that is also very close. We considered the weighted mean between $P_\mathrm{cyc,P}$ and the value from B18 as a final estimate. HD~155886 has $P_\mathrm{cyc,H}$ and $P_\mathrm{cyc,QP}$ from O18, and an estimate from B18, which are all substantially different from one another. We removed this star to avoid ambiguities. HD~1835 and HD~20630 have no $P_\mathrm{cyc}$ measurement from O18 and they have each two different estimates from B18. We removed these targets to avoid ambiguities. HD~190007 has $P_\mathrm{cyc,P}$ and $P_\mathrm{cyc,QP}$ from O18 that are almost identical, while a substantially different one comes from B18. We considered $P_\mathrm{cyc,QP}$ from O18 as the final one. HD~190406 has $P_\mathrm{cyc,QP}$ from O18 that is very close to one of the two estimates provided by B18. For the final value, we computed the weighted mean of the two. Despite HD~201092 has $P_\mathrm{cyc,P}$ from O18, which is very close to the value provided by B18, we discarded this star from our sample as its determination of the activity cycle period appears rather uncertain due to the pronounced caothicity of its  activity indicators time-series \citep{61CygB06,61CygB12}.  HD~76151 and HD~78366 both have only $P_\mathrm{cyc,H}$ from O18 and an estimate from B18 that is very close. The final estimate is taken as a weighted mean of the two values. HD~82443 has $P_\mathrm{cyc,P}$ and $P_\mathrm{cyc,QP}$ from O18 that are not in agreement, and another value substantially different from these two from B18. We removed this target to avoid ambiguities. The final uncertainties on $P_\mathrm{cyc}$ are computed following a standard error propagation from those listed by the two catalogs.

In relation to $P_\mathrm{rot}$, to obtain a final estimate we computed a simple arithmetic mean between the two catalog estimates, except for the stars HD~26923 and HD~32147, which were excluded from our final catalog because of the pronounced disagreement between the two measurements ($> 30$ \,\%). Moreover, we computed values for the chromospheric activity index $\log_{10} R^{'}_\mathrm{HK}$ by taking the arithmetic mean of the measurements available from O18 and B18 for those stars that are in common. For this parameter we find an excellent agreement (well within 10\,\%) between the two catalogs.

By relying on a calibration of the Rossby number (Ro) provided by \cite{Corsaro21} using the $(B-V)$ color index, we find that our stellar sample has a mean $\mbox{Ro} = 0.523 \pm 0.481$. This shows that, albeit our catalog comprises stars spanning from K to F spectral types and from main sequence to early giants, their Rossby number is on average close to that found in the Sun (Ro$_{\odot} = 0.496$), meaning that it is reasonable to study the sample as a whole. 

\subsection{Stellar metallicity and Age}
The period measurements were complemented with additional measurements of stellar metallicity and age by cross-matching our compiled catalog with the newest available ESA Gaia \citep{Gaia} DR3 catalog of additional astrophysical parameters from the Apsis processisng chain (\texttt{paramsup}, \citealt{GaiaDR3Apsis}), and with high-resolution spectroscopy measurements. 

The Gaia DR3 catalog provides metallicities $\left[\mbox{M/H}\right]$ from low-resolution spectra for 47 stars of our sample. For the high-resolution measurements instead, we considred literature values from works exploiting a consistent data analysis procedure, as obtained by \cite{VF05} (31 stars), \cite{Brewer16} (20 stars), \cite{Luck15} (6 stars), and \cite{Luck17} (39 stars). Multiple high-resolution $\mh$ measurements for an individual star, when available, are combined through an arithmetic mean.

Stellar ages are obtained to large extent from the Gaia DR3 catalog, which provides isochrone-based estimates (related to BaSTI stellar evolution models, \citealt{BaSTI18}) for 33 stars of our sample. For HD~146233 (18 Sco) we replaced the Gaia DR3 estimate with an age based on asteroseismology \citep{18Sco12}. Additional measurements are taken from the \texttt{StarHorse} Gaia EDR3 catalog of astrophysical parameters \citep{StarHorseEDR3}. In this catalog we found ages for the stars HD~152391, HD~166620, HD~219834A, HD~219834B, HD~26913, HD~21749, HD~154577, HD~75332, HD~88373, which were instead missing in the Apsis Gaia DR3 catalog. We complete our set of ages by including measurements for the Sun \citep{Bonanno15}, and the well known stars  HD~124897 \citep[Arcturus,][]{Arcturus11} and HD~201091 \citep[61 Cyg A,][]{61Cyg08}, thus leading to a total of 45 stars having an age estimate. All of the collected characteristics for the ensemble is provided in Table~\ref{table1} in a machine-readable format.

\begin{deluxetable*}{lcl}
\tablecaption{Metadata for Compiled catalog \label{table1}}
\tablewidth{0pt}
\tablecolumns{3}
\tablehead{
\colhead{Unit} &
\colhead{Label} &
\colhead{Description}
}
\startdata
---         &   Name       & Star Name, HD catalog or Sun \\
---         &   DR3        & Gaia DR3 identifier \\
d           &   Prot       & Rotational period, days \\
yr          &   Pcyc       & Activity cycle, years\\
yr          & e\_Pcyc      & Standard deviation of Pcyc \\
dex         &   [M/H]-LR & Gaia DR3 low-resolution-based metallicity \\
dex         &   [M/H]-HR & High-resolution (literature) metallicity \\
dex         &   [M/H]    & Final metallicity \\
Gyr         &   Age        & Age, Gigayears \\
dex         &   logRHK     & chromospheric activity index, $\log_{10} R^{'}_\mathrm{HK}$ \\
M$_\mathrm{\odot}$    &   Mstar      & Mass of star \\
dex         &   logg       & log, surface gravity, $\log g$ \\
L$_\mathrm{\odot}$    &   Lstar      & Luminosity of star \\
mag         &   B-V        & B-V color \\
d           &   tau        & Convective turnover time from \citealt{Corsaro21}, days \\
\enddata
\end{deluxetable*}

\section{Bayesian analysis}
\label{sec:bayes}
\subsection{Models}
\label{sec:models}
As originally presented by \cite{Saar99} and then further discussed by e.g. \cite{Brandenburg17} (see also O18), the magnetic activity cycles are distributed along two potentially different regimes that have been termed active and inactive branches. In this work we focus on the observable quantities $\omega_\mathrm{cyc}$ and $\Omega$, and following Eq.~(\ref{eq:cyc_rot}) we assume that these two observables are related by a simple power-law relation of the type
\begin{equation}
\omega_\mathrm{cyc} = \beta  \Omega ^{\nu} \, .
\end{equation}
We therefore set our analysis in the plane $\log_{10} \ocy $ -- $\log_{10} \Omega$ plane, hereafter OCOR (Omega Cycle versus Omega Rotation) plane for simplicity, where the power-law relation can be linearized. Following the literature, at first place we consider two regimes that can be modeled through the presence of two simultaneous linear relations of the type
\begin{equation}
\log_{10} \omega_\mathrm{cyc} = \nu_1 \log_{10} \Omega + \log_{10} \beta_1
\label{eq:line_1}
\end{equation}
and
\begin{equation}
\log_{10} \omega_\mathrm{cyc} = \nu_2 \log_{10} \Omega + \log_{10} \beta_2 \, ,
\label{eq:line_2}
\end{equation}
which we term model $\mathcal{M}_1$. In the second case we consider 
\begin{equation}
\log_{10} \omega_\mathrm{cyc} = \nu_0 \log_{10} \Omega + \log_{10} \beta_0
\end{equation}
which we term model $\mathcal{M}_2$. This second model has the purpose of allowing us to verify whether a single power law, hence a single regime in the evolution of the magnetic activity cycle, ought to be preferred over two separate ones.

In all cases here presented, the coefficients $(\nu_i,\beta_i)$ for $i=0,1,2$, will be estimated through a statistical inference (see Sect.~\ref{sec:likelihood} for details). Besides, a model comparison process will allow us to draw conclusions on whether one or two regimes ought to be preferred in the light of the existing data (Sect.~\ref{sec:inference}). 


\subsection{Likelihood}
\label{sec:likelihood}
For the simultaneous fitting of two regimes as presented through model $\mathcal{M}_1$ we do not perform an a priori division of the data sample in two parts, which would otherwise result in imposing a somewhat arbitrary cut. We instead automatically identify the presence of any internal partition by means of a purely statistical approach. For this purpose we consider a Gaussian mixture likelihood, consisting of two components of the type $y^{(1)} = \nu_1 x + \log_{10} \beta_1$, $y^{(2)} = \nu_2 x + \log_{10} \beta_2$, with $\nu_1,\nu_2$ the two line slopes, and $\log_{10} \beta_1,\log_{10} \beta_2$ the two line offsets. The corresponding likelihood can be expressed as
\begin{equation}
{\mathcal{L}} 
\left(\boldsymbol{\theta}\right) = 
\prod_{i=1}^N \left[ A_i \left(\boldsymbol{\theta}\right) + B_i \left(\boldsymbol{\theta}\right) \right] \, ,
\end{equation}
where 
\begin{equation}
A_i \left(\boldsymbol{\theta}\right) = \frac{\sin^2 \psi}{\sqrt{2 \pi} \sigma_i} 
\exp \left[ -\frac{\left(y_i - \nu_1 x_i - \log_{10} \beta_1 \right)^2}{2 \sigma^2_i} \right]
\end{equation}
and
\begin{equation}
B_i \left(\boldsymbol{\theta}\right) = \frac{\cos^2 \psi}{\sqrt{2 \pi} \sigma_i} 
\exp \left[ -\frac{\left(y_i - \nu_2 x_i - \log_{10} \beta_2 \right)^2}{2 \sigma^2_i} \right]
\end{equation}
are the two components of the Gaussian mixture, with $\boldsymbol{\theta} = (\nu_1,\beta_1,\nu_2,\beta_2,\psi)$  the parameter vector containing the free parameters that need to be fitted, and $\sigma_i$ the uncertainties on the dependent variable $y_i$. We therefore have to estimate five free parameters in the case of two simultaneous regimes, $\psi$ being the parameter that controls the relative weight of each of the two components of the mixture. In the way we presented the models, we further assume that the measurements $\omega_{cyc}$ are log-normally distributed \citep[e.g. see][]{Corsaro13,Bonanno14}. This is because  the residuals arising from the difference between the observed and predicted logarithms of the activity cycle rate are assumed to be Gaussian distributed.

For avoiding numerical overflows, which may arise from the exponential term in the likelihood, we switch to the log-likelihood and rely on the identity
\begin{eqnarray}
    &&\ln(A+B) = \ln\left[\exp{\left(\ln A-\ln B\right)}+1 \right] + \ln B \, .
\end{eqnarray}
Therefore, we compute each term $i$ using two different expressions. By defining the quantity
\begin{equation}
\Delta_\mathrm{ij} \equiv - \frac{1}{2 \sigma_i^2} \left[ y_i - y^{(j)}_i \right]^2
\end{equation}
we obtain for $\Delta_\mathrm{i1} \geq \Delta_\mathrm{i2}$ \, ,
\begin{equation}
\begin{split}
\Lambda^{(a)}_i \left( \boldsymbol{\theta} \right) =& \ln \left( \frac{\sin^2 \psi}{\sqrt{2 \pi} \sigma_i} \right)  + \Delta_\mathrm{i1} +\\
&\ln \left[ 1 + \cot^2 \psi \exp{\left( \Delta_\mathrm{i2} - \Delta_\mathrm{i1}\right)} \right] \, ,
\end{split}
\end{equation}
and vice versa for $\Delta_\mathrm{i1} < \Delta_\mathrm{i2}$
\begin{equation}
\begin{split}
\Lambda^{(b)}_i \left( \boldsymbol{\theta} \right) =& \ln \left( \frac{\cos^2 \psi}{\sqrt{2 \pi} \sigma_i} \right) + \Delta_\mathrm{i2} + \\
&\ln \left[ 1 + \tan^2 \psi
\exp{\left( \Delta_\mathrm{i1} - \Delta_\mathrm{i2}\right)} \right] \, . 
\end{split}
\end{equation}
In this way we can compute a final log-likelihood as
\begin{equation}
\Lambda \left( \boldsymbol{\theta} \right) = \sum_{i=1}^N \Lambda_i^{(a,b)} \left( \boldsymbol{\theta} \right) \, ,
\end{equation}
where $\Lambda^{(a,b)}$ can be either $\Lambda^{(a)}$ or $\Lambda^{(b)}$ depending on the conditions defined above. 

For treating the case of model $\mathcal{M}_2$, consisting of a single linear relation, we instead rely on a standard Gaussian likelihood. We note that the Gaussian mixture likelihood here  presented and often adopted in Bayesian cluster analysis is a more generalized form of the standard Gaussian likelihood. The two likelihood functions become identical by imposing $\psi = 0$ or $\pi/2$, i.e. by assuming that the sample is not internally split in two separate subsets.

\subsection{Bayesian inference and model comparison}
\label{sec:inference}
By means of the log-likelihood function presented in Sect.~\ref{sec:likelihood}, we perform a Bayesian inference by adopting uniform prior distributions for each free parameter involved in the fitting process. We note that although the free parameter $\psi$ is not directly appearing in the model equations, it is an important fitting parameter because it controls how the data sample is clustered in the light of the two components of the mixture. Given that $\psi$ has to satisfy the identity $\sin^2 \psi + \cos^2 \psi = 1$, we have adopted a uniform prior in the range $[0,\pi/2]$ to allow for all the possible outcomes. We further point out that for considering more reliable uncertainties on $\ocy$ than those provided in the literature --- which are, in some cases, unrealistically more precise than what can be obtained for our Sun --- we have inflated them for our analysis by incorporating a systematic term due to the intrinsically
stochastic character of the dynamo mechanism \citep{2000PhRvL..85.5476M}.
We can estimate this term by assuming it is of the same order of the dispersion of the solar activity cycle 
period. For computing a reliable estimate, we consider the longest available set of solar activity cycle maxima, as obtained from sunspots time-series covering a total of 18 different activity cycles (in the years from 1818 to 2014)\footnote{Data are taken from the SILSO data/image, Royal Observatory of Belgium, Brussels, https://www.sidc.be/silso/ .}. The result yields $\sigma^\mathrm{sys}_{P_\mathrm{cyc,\odot}} = 0.24\,$yr.

To perform the Bayesian parameter estimation we make use of the public tool \textsc{D\large{iamonds}}\footnote{https://github.com/EnricoCorsaro/DIAMONDS} \citep{Corsaro14}, which exploits a nested sampling Monte Carlo algorithm \citep{Skilling04}. The results of our analysis are presented in Sect.~\ref{sec:results}. In addition to the parameter estimation problem we perform a Bayesian model comparison through the odds ratio, which for two competing models $\mathcal{M}_i$ and $\mathcal{M}_j$ is defined as 
\begin{equation}
\mathcal{O}_{ij} = \frac{\mathcal{E}_i}{\mathcal{E}_j} \frac{\pi \left(\mathcal{M}_i \right)}{\pi \left(\mathcal{M}_j \right)} \, .
\label{eq:odds}
\end{equation}
In our application, we can discard the term containing the ratio of the model priors $\pi \left(\mathcal{M}_i\right)$, which we assume to be the same for each of the models considered, so that the odds ratio is entirely based on the evaluation of the Bayesian evidences $\mathcal{E}_i$. We refer to this odds ratio as the Bayes factor, which we can express in logarithmic form as $\ln \mathcal{B}_{ij} = \ln \mathcal{E}_i - \ln \mathcal{E}_j$. According to the Jeffreys' scale of strength \citep{Trotta08}, we are therefore able to understand whether a given model could be favored or not against its competitor. The Bayesian evidences are computed for the pair of models $(\mathcal{M}_1$, $\mathcal{M}_2)$ and are provided as a direct output of the computation by \textsc{D\large{iamonds}}. We note that the evaluation of Eq.~(\ref{eq:odds}) is not based on any approximation such as that of the Bayesian Information Criterion, but on the full numerical calculation of the Bayesian evidence as obtained by \textsc{D\large{iamonds}} from the sampling of the multi-dimensional posterior probability distributions.

\section{Results}
\label{sec:results}
\begin{figure}
    \centering
    \includegraphics[width=8.5cm]{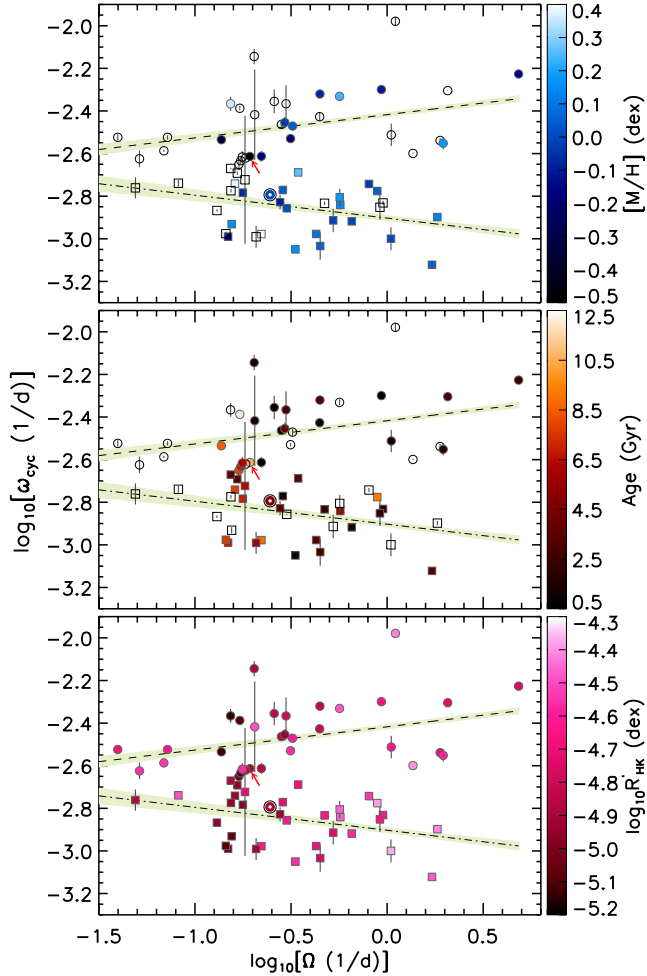}
    \caption{Magnetic activity cycle rate $\ocy$ as a function of the rotation rate $\Omega$. The dashed and dot-dashed lines represent the fits obtained from the double linear regression (Eqs.~\ref{eq:line_1} and \ref{eq:line_2}, respectively). The symbol type represents the group classification as stemming from the fits (circles for Group 1 and squares for Group 2). Open symbols are stars for which the color-coded measurement was not available. 1-$\sigma$ uncertainties on $\omega_\mathrm{cyc}$ as well as the 3-$\sigma$ credible region of each fit are overlaid. The red arrow shows the star HD~103095, having  $\left[\mbox{M/H}\right] \simeq -1.34$, significantly lower than the average metallicity of the sample. The available measurements of $\left[\mbox{M/H}\right]$, Age, and $\log_{10} R^{'}_\mathrm{HK}$, are shown as color-coded information in the \textit{top}, \textit{middle}, and \textit{bottom panels}, respectively. The Sun with its symbol is indicated and included in G2, color-coded according to the corresponding measurements.}
    \label{fig:ocor}
\end{figure}

The output of the model comparison process yields a clear preference for the model that incorporates two different regimes in the OCOR plane, namely favoring model $\mathcal{M}_1$ with $\ln \mathcal{B}_{1,2} \simeq 2.6 \cdot 10^4$, largely beyond a strong evidence condition. The result of the fit is shown in Fig.~\ref{fig:ocor}, while the estimates of the free parameters of model $\mathcal{M}_1$ are $\overline{\nu}_1 = 0.109_{-0.005}^{+0.005}$, $\log_{10} \overline{\beta}_1 = -2.417^{+0.004}_{-0.004}$, $\overline{\nu}_2 = -0.108_{-0.008}^{+0.007}$, $\log_{10} \overline{\beta}_2 = -2.903^{+0.005}_{-0.005}$. We obtain a weight parameter $\overline{\psi} = 0.69^{+0.07}_{-0.06}$, showing that in the mixture both groups are playing equally important roles  ($\sin^2 \overline{\psi} \simeq 0.49$). 

We classify the stars in two groups according to their relative difference in $\log_{10} \ocy$ with respect to each line of the fit and identify stars having a larger $\ocy$ as Group 1 (G1) stars, and stars having a lower $\ocy$ as Group 2 (G2) stars. The Sun clearly belongs to G2 as it is well in line with the regime indicated by Eq.~(\ref{eq:line_2}).

\begin{figure}
   \centering
   \includegraphics[width=8.5cm]{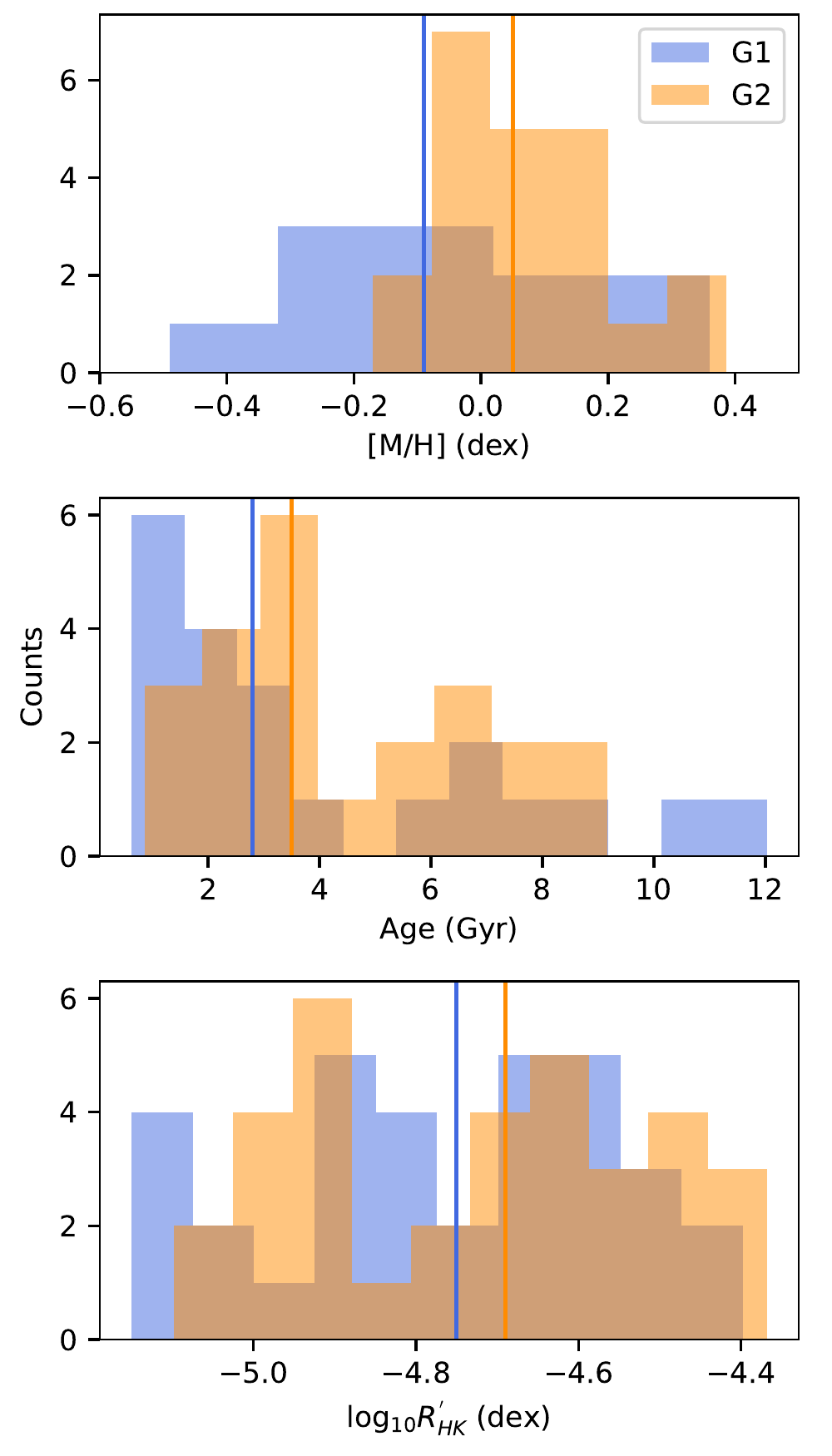}
    \caption{Distribution of metallicity $\left[\mbox{M/H}\right]$, Age, and chromoshperic activity index $\log_{10} R^{'}_\mathrm{HK}$ from top to bottom for the stars of Group 1 and 2 as identified in Sect.~\ref{sec:results}. The overlaid vertical lines, having the same color as the histograms, represent the corresponding median values for each group. The metallicity range has been restricted to a range similar to that adopted in Fig.~\ref{fig:ocor} for clarity purposes.}
    \label{fig:hist}
\end{figure}

\section{Discussion}
\label{sec:discussion}
In order to understand the potential origin of the dichotomy shown in Fig.~\ref{fig:ocor} we analyze it in relation to the fundamental stellar properties of metallicity $\left[\mbox{M/H}\right]$, Age, and chromospheric activity index $\log R^{'}_\mathrm{HK}$ for each of the two groups identified.

In relation to metallicity, we find a significant correlation (Pearson's correlation coefficient $\rho = 0.78$) between the low-resolution values obtained from Gaia DR3 and the high-resolution ones, for the stars that are in common (a total of 33). We therefore combine the two sets by first shifting the zero point of the Gaia $\mh$ set to match that of the high-resolution $\mh$ set (-0.185 dex as obtained from a linear regression over the two sets), and by subsequently computing an arithmetic mean between the two. The final metallicity sample therefore represents stars having both Gaia DR3 and high-resolution measurements available. The presence of an offset in the Gaia metallicities may have an impact on the adopted stellar ages. We therefore caution the reader that this represents a potential limitation of our work in relation to the interpretation of the age distributions.

To highlight any difference between Group 1 and 2 we additionally show their corresponding distributions as a function of stellar properties in Fig.~\ref{fig:hist}. We find that the stars in G2 are about 40\,\% more metal rich than the stars in G1, with median metallicity values $\left[\mbox{M/H}\right]_\mathrm{G1} \simeq -0.09\,$dex and $\left[\mbox{M/H}\right]_\mathrm{G2} \simeq +0.05\,$dex. According to a Kolmogorov-Smirnov test, the two metallicity distributions appear different to a p-value level of 0.015, close to the usual limit adopted for a strong detection (which is set to 0.003). Interestingly, we find that the stars in G2 are about 25\,\% older than those in G1, having median values Age$_\mathrm{G1} \simeq 2.8\,$Gyr and Age$_\mathrm{G2} \simeq 3.5\,$Gyr. Although the age difference between the two groups cannot be considered statistically significant (the associated p-value of the two distributions is 0.17), the trend appears in line with that found for metallicity because stars with lower metallicity have a reduced opacity, which leads to stronger radiative fluxes, hence a faster evolution. For the sake of completeness, we also performed a comparison with ages computed from gyrochronology \citep{Barnes07}. Here we note that while ages from gyrochronology are in agreement with those from isochrones for G1 (where we obtain a median gyro-age of 2.6 Gyr), there is a discrepancy for stars in G2, with gyro-ages pointing to a median of 1.6 Gyr. According to gyrochronology, this would imply that stars in G2 should rotate slower than observed to match the isochrone ages. While the investigation of this difference is beyond the scope of this work, we cannot exclude that stars in G2 may be subject to a different gyrochronological calibration than the one presented by \cite{Barnes07}.

In relation to the activity index $\log_{10} R^{'}_\mathrm{HK}$ instead, we find two rather similar median values, $\log_{10} R^{'}_\mathrm{HK, G1} \simeq -4.75\,$dex and $\log_{10} R^{'}_\mathrm{HK, G2} \simeq -4.69\,$dex. Here one could in principle attempt to rely on an age-activity relationship such as the one proposed by \cite{Mamajek08} to estimate an age difference between the two groups from the choromspheric activity level alone. However, we point out that according to a Kolmogorov-Smirnov test the two distributions in $\log_{10} R^{'}_\mathrm{HK}$ are essentially indistinguishable (p-value $> 0.9$). This implies that the variance in chromospheric activity between G1 and G2 is just too small to be reliably used to infer any dissimilarity in age. Besides, the general differences identified by our computation of the median estimators appear clearly visible in the distributions shown in Fig.~\ref{fig:hist} in relation to $\mh$. 

We note that we have also inspected available measurements for mass, $\log g$, and luminosity of the stars belonging to G1 and G2, and found that while the median values for mass and $\log g$ of the two groups are very similar, pointing to $M \simeq 1\,M_{\odot}$ and $\log g \simeq 4.44\,$dex, there is a $\sim 14$\,\% difference in luminosity (median values $L_\mathrm{G1} \simeq 0.87\,L_\mathrm{\odot}$ and $L_\mathrm{G2} \simeq 0.76\,L_\mathrm{\odot}$). This not only suggests that is the metallicity, and not the stellar mass, that could be playing a role in the different dynamo mechanism in action between the two groups, but also that there is a qualitative agreement among the trends observed in $\mh$, Age, and luminosity.

\section{Conclusions}
\label{sec:conclusions}
In this study we have shown that stars can be classified into two groups according to whether the period of the activity cycle decreases or increases with rotation. In both cases the simple scaling law represented in Eq.~(\ref{eq:cyc_rot}) seems to hold for a wide range of rotations.  Albeit 
the average rotation rate of the two groups is nearly the same, cycle periods for the group G1 are about a factor two shorter than for the group G2.  We suggest that this dichotomy is likely originating from a different chemical composition of the stars. Finally, we would like to stress that the Sun does not occupy any special position in this context.

Can these two scaling laws be understood in terms of dynamo theory? As discussed in the introduction the positive slope can be explained already in terms of simple kinematic dynamo theory. The negative slope does not have this interpretation. However if we assume that the dynamo in the Sun is ruled by the meridional circulation at the bottom of the convection zone, then a higher rotation rate would imply weaker meridional circulation (by angular momentum conservation). This would produce a smaller cycle period because the cycle period is determined by the strength of the meridional circulation \citep{2001ASPC..248..235D,2013GApFD.107...11B}.
What is the role of metallicity in this context? Stars with reduced metallicity have higher luminosity, therefore higher eddy diffusivity, at least according to mixing-length theory. On the contrary, a smaller luminosity 
implies a reduced diffusivity and therefore a stronger Reynolds number of the meridional circulation for this type of dynamo action. 

We plan to further explore these arguments in detail in a future work that will also exploit a larger set of metallicity measurements as obtained by the soon to come ESA PLATO mission \citep{Rauer14PLATO}.

\acknowledgements{We thank Gustavo Guerrero for comments on the manuscript, Arman Khalatyan, Hans-Erich Froehlich, Antonino Lanza, Sergio Messina, Santi Cassisi and Isabella Pagano for helpful dicussions. A special thanks to the anonymous referee who has helped us in greatly improving the manuscript. A.B. and E.C. acknowledge support from PLATO ASI-INAF agreement no. 2015-019-R.1-2018.}


\end{document}